\catcode`@=11 
\def\binrel@#1{\setbox\z@\hbox{\thinmuskip0mu
 \medmuskip-1mu\thickmuskip\@ne mu$#1\m@th$}%
 \setbox\@ne\hbox{\thinmuskip0mu\medmuskip-1mu\thickmuskip
 \@ne mu${}#1{}\m@th$}%
 \setbox\tw@\hbox{\hskip\wd\@ne\hskip-\wd\z@}}
\def\binrel@@#1{\ifdim\wd2<\z@\mathbin{#1}\else\ifdim\wd\tw@>\z@
 \mathrel{#1}\else{#1}\fi\fi}
\def\pmb{\let\next\relax\ifmmode\def\next{\mathpalette\pmb@}\else
 \let\next\pmb@@\fi\next}
\def\pmb@@#1{\leavevmode\setbox\z@\hbox{#1}\kern-.025em\copy\z@\kern-\wd\z@
 \kern-.05em\copy\z@\kern-\wd\z@\kern-.025em\raise.0433em\box\z@}
\newdimen\pmbraise@
\def\pmb@#1#2{\setbox\thr@@\hbox{$\m@th#1{#2}$}%
 \setbox4\hbox{$\m@th#1\mkern.7794mu$}\pmbraise@\wd4
 \divide\pmbraise@18
 \binrel@{#2}\binrel@@{\mkern-.45mu\copy\thr@@\kern-\wd\thr@@
 \mkern-.9mu\copy\thr@@\kern-\wd\thr@@\mkern-.45mu\raise\pmbraise@\box\thr@@}}
\catcode`@=12 

\def\lsim{\mathrel{\rlap{\lower4pt\hbox{\hskip1pt$\sim$}}
    \raise1pt\hbox{$<$}}}         
\def\gsim{\mathrel{\rlap{\lower4pt\hbox{\hskip1pt$\sim$}}
    \raise1pt\hbox{$>$}}}         

\def\overleftrightarrow#1{\vbox{\ialign{##\crcr
    $\leftrightarrow$\crcr
    \noalign{\kern 1pt\nointerlineskip}
    $\hfil\displaystyle{#1}\hfil$\crcr}}}
    \long\def\caption#1#2{{\setbox1=\hbox{#1\quad}\hbox{\copy1%
    \vtop{\advance\hsize by -\wd1 \noindent #2}}}}

\def\frac#1/#2 {{\textstyle {#1\over #2}}}
\def\b{\vec}

\def\vr{\vec r\,}
\def\s0{\sigma_0}
\def\si{\sigma}
\def\br{<\!}
\def\ke{\!>}

\def\a{\alpha}
\def\be{\beta}
\def\D{\Delta}
\def\d{\delta}
\def\e{\epsilon}

\def\underrightarrow#1{\vtop{\ialign{##\crcr
    $\hfil\displaystyle{#1}\hfil$\crcr\noalign{\kern0pt\nointerlineskip}
     \rightarrowfill\crcr\noalign{\kern0pt}}}}

\def\be{\begin{equation}}
\def\eq{\end{equation}}
\def\pb{\bar p}
\documentstyle[preprint,revtex]{aps}
\begin{document}
RevTex
\vskip 9pt
\begin{title}
SEMI-CLASSICAL DESCRIPTION OF\\
ANTIPROTON CAPTURE ON ATOMIC HELIUM
\end{title}
\author{W. A. Beck$^{(1),(2)}$, L. Wilets$^{(1)}$ and M. A.
Alberg$^{(1),(3)}$}
\begin{instit}
$^{(1)}$Department of Physics, University of Washington, Seattle, WA 98195\\
$^{(2)}$ Quantum Medical Systems, Issaquah, WA 98027\\
$^{(3)}$Department of Physics, Seattle University, Seattle, WA 98122
\end{instit}

\begin{abstract}

A semi-classical, many-body atomic model incorporating a momentum-dependent
Heisenberg core to stabilize atomic electrons is used to study antiproton
capture on Helium.  Details of the antiproton collisions leading to
eventual capture are presented, including the energy and angular momentum
states of incident antiprotons which result in capture via single or double
electron ionization, i.e. into [He$^{++}\,\bar p$ or He$^{+}\,\bar p$], and
the distribution of energy and angular momentum states following the Auger
cascade.  These final states are discussed in light of recently reported,
anomalously long-lived antiproton states observed in liquid He.

\end{abstract}

\section{INTRODUCTION}

The ``trapping" of antiprotons stopped in liquid He into long-lived states
has been reported recently by Iwasaki {\it et al.}\cite{iwa}  Approximately
3.6\%  of the incident antiprotons exhibit delayed decay, of the order of
microseconds compared with picoseconds for prompt decays, opening
possibilities for further experimentation, including antihydrogen production.

The initial slowing and capture of in-flight antiprotons is via energy loss
due to Auger excitation and ionization of the atomic electrons, while the later
stages are dominated by radiation, and eventual annihilation.  Auger
transitions are characteristically many orders of magnitude more rapid than
radiative decay, making the dynamics of the Auger process central to
understanding the delayed antiproton annihilation.

Yamazaki {\it et al.}\cite{yam} , following the work of Condo\cite{con}
 and Russell\cite{rus},
proposed that the long-lived antiprotons were captured into ``meta-stable,
circular states," of principal quantum number $n_0 = \sqrt{M^*/m_e} \approx
38$ and with $\ell_0 \approx n_0-1$, of the exotic neutral He$^+\,\bar p$
atom.  ($M^*=\frac4/5 M_{\bar p}$ is the antiproton reduced mass with
respect to He and $m_e$ is the electron mass.)

{}From these states further Auger transitions would be highly inhibited:

1) $\D \ell>1$ is highly unfavored and the atomic excitation energies are
large compared with the antiproton spacing for $\D n=\D\ell=-1$, and

2) The Stark effect, which normally admixes $\ell=0$ states due to the
presence of other atoms in the liquid and enhances $\pb$ annihilation in
the nucleus for atoms stripped of inner electrons, is suppressed due to the
removal of $\ell$-degeneracy by the presence of the remaining electron.

Thus decay is limited to the much slower radiative transitions, which have
small photon energy, and antiproton annihilation is delayed.

In order to study the capture of antiprotons by He atoms, we modeled
antiproton collisions with He using a classical description of the
antiprotons and a semiclassical description of the electrons, in which the
electrons of the target system are stabilized by a momentum-dependent
Heisenberg core[5]; the Pauli principle can be ignored here since
the atomic electrons are in antiparallel states.

\section{QUANTUM AND ADIABATIC SOLUTIONS of the He-$\pb$ system}

Consider now the quantum states of this system, an exotic ``molecule"
consisting of the He nucleus, i.e. an $\a$ particle, the antiproton and
$N=$0, 1, or 2 electrons.  Working in atomic units, with $\hbar=m_e=e=1$
(ignoring the negligible effect of electron reduced mass), the Hamiltonian
is given by

\begin{equation}
H = {P^2\over 2M^*} - {2\over R} + \sum_{i=1}^N \left[ {p_i^2\over 2} -
 {2\over r_i}+{ 1\over |\b R - \b r_i| } \right]+{1\over r_{12}}\d_{N,2} \,,
\end{equation}
where $\b r_i$ and $\b R$ are the electron and antiproton coordinates
relative to the helium nucleus, and $\b p_i$ and $\b P$ are the
corresponding momenta; $M^* =1469$ is the reduced mass of the antiproton
in atomic units.

The zero electron (double ionization) problem is simply the He$^+$ ion with
the $\bar p$ replacing the electron.  The energy levels are given by
\be
E_n = - {Z^2 M^*\over 2 n^2} = -{2 M^* \over n^2}\,. \label{eq:he}
\eq
Note that $n_0 =\sqrt{M^*}\approx 38$ corresponds to a $\pb$ orbit of the
same size and energy as the 1s electron orbit; because of the large $M^*$
and consequent large quantum numbers for the antiproton states of interest,
the antiproton may be treated classically.

The one electron (single ionization) problem is described well in the
adiabatic Born-Oppenheimer approximation.  We note that this two-center,
one-electron problem is separable in prolate spheroidal coordinates and can
thus be calculated to high accuracy, cf \cite{m&f}, \cite{shim}. In terms of
the coordinates defined above, the energy levels $\epsilon_n(R)$ of one
electron in the potential of the He nucleus and fixed antiproton are given by
\be
\left[{p^2\over 2} -{2\over R} -{2\over r}+ {1\over|\b R - \vr|}\right]
\psi_n(\vr, \b R) = \epsilon_n(R) \psi_n(\vr,\b R)      \label{eq:BO}
\eq
and are the potential energy for the Born-Oppenheimer antiproton
eigenvalue equation,
\be
\left[ {P^2\over 2M^*} + \epsilon_n(R)  \right] \Phi_{n,m}(\b R) =
E_{n,m} \Phi_{n,m}(\b R)\,.
\eq
Here $P^2=-R^{-1}(\partial ^2/\partial R^2) R + L^2/R^2$ includes vibrational
and rotational excitation.
These Born-Oppenheimer states are not stationary states of the full problem,
however, since there remain velocity terms in the full Schroedinger equation
which couple the adiabatic levels.  Quite generally we can expand the total
wave function in terms of the eigenfunctions of Eq.~(\ref{eq:BO}).

\be \Psi(\vr,\b R)=\sum_n\Phi_{n}(\b R)\psi_n(\vr,\b R)\,.\eq

The Schroedinger equation for the $\Phi_n$ is then
\be
\left[{P^2\over 2M^*} +\epsilon_n(R)-E\right]\Phi_n+{1\over 2M^*}
\sum_{n'}\left[2\br n|\b P|n'\ke\cdot \b P+\br n|P^2|n'\ke\right]\Phi_{n'}=0
\eq
where $|n\ke$ corresponds to $\psi_n$ and the bra-ket integration is over
$\vr$.

In a time-dependent calculation, these coupling terms lead to Auger
transitions between the adiabatic states.  Note that the first coupling term
is essentially of the dipole form (the operator is $\b P$).  Thus
transitions with $|\D \ell|>1$ are (progressively more) inhibited.

Alternately, Yamazaki and Ohtsuki \cite{yamoh} have obtained approximate
solutions to the Born-\~Oppenheimer problem using configuration-mixing
techniques, in which antiproton states are calculated with the electron in
the ground state, and then mixed with excited states of the electron to
obtain the system levels shown in Fig. 1.  Here, the system levels
$L = N-1$ are proposed as the boundary of the allowed $E, L$ states, with
the states with $L > 31$ proposed as the region of metastability.

Because of the infinite range of the central Coulomb potential, the Born-
Oppenheimer energy level diagram of $E(L)$, where $L$ is the total angular
momentum, in theory extends to the right indefinitely; this can be seen by
considering first a simple $\a\pb$ state in the energy region covered in
Fig. 1.  An electron can then be placed in a hydrogenic orbit of arbitrarily
small energy and arbitrarily large angular momentum without disturbing the
$\pb$. Only some of these states were observed in our calculations, however;
electrons excited to higher angular momentum states during the collision
process also gained enough kinetic energy to escape from the (exotic) atomic
system.

Capture of $\pb$ into the two electron negative ion was also not observed
in our calculations; this negatively charged system appeared only briefly
during the initial stages of antiproton capture, and was unstable to
electron ionization during the antiproton capture and decay process.

\section{THE SEMI-CLASSICAL MODEL}
\label{sec:mod}
In the semi-classical model  \cite{kir}, the Hamiltonian of the undisturbed
Helium atom is
\be
H_{sc} = {p_1^2\over 2} +{p_2^2\over 2} -{2\over r_1} - {2\over r_2}
+ {1\over r_{12}} + V_H\,
\eq
with
$V_H$ the momentum-dependent Heisenberg core which prevents the collapse of
the electrons into the He nucleus, given by

 \be V_H = {\xi_H\over 4 \a r^2}\exp\big\{ \a [1 - (r\,p/\xi_H)^4]\big\} \eq

We choose \cite{kir} $\a = 1.0,\ \xi_H = $2.767. The Hamiltonian then minimizes
at an energy of $-2.78$ with $r_1= r_2 = 0.63$, in fair agreement with the
exact ground state energy of $-2.9037\cdots\ $  at an electron mean radius
of $<r>\approx 0.59$.

The total Hamiltonian is then
\be
H_{sc} = {P^2\over 2M^*} - {2\over R} + \sum_{i=1}^2 \left[ {p_i^2\over 2} -
 {2\over r_i}+{ 1\over{|\b R - \vr_i|} } +V_H(r_i, p_i)\right]+{1\over r_{12}}
\eq

To model collisions, Hamilton's classical equations of motion
 \be
 {dx_i\over dt}={\partial H_{sc}\over \partial p_i}\,,\qquad
 {dp_i\over dt}=-{\partial H_{sc}\over \partial x_i}
 \eq
are solved for $\vr_1,\ \b p_1,\ \vr_2,\ \b p_2,\ \b R,\ \b P\,$, the
coordinates describing our semi-classical system, and integrated over time
using the venerable ordinary differential equation routine ODE.

\section{COLLISION CALCULATIONS}
\label{sec:calc}
Monte Carlo calculations of antiproton collisions with He were performed
as follows:

(1) The target He atom was prepared in its ground state with random
orientation and parity inversion of the electron coordinates and momenta, as
described in \cite{wil2}.

(2) The projectile $\bar p$ was launched at this target atom with an initial
energy of about 80 eV (about 3.0 a.u.) and with an impact parameter $b$
randomized with equal areas ($\pi db^2$) up to an energy-dependent $b_{max}$.

(3) Sequences of collisions were followed from one encounter to the next
until capture occurred.

Our initial energy and $b_{max}$ were chosen so that collisions would never
result in antiproton capture on the first encounter, and the final energy of
the antiproton after each collision could be used as the starting energy for
a subsequent collision, again with Monte Carlo initial conditions for the
ground state of the new target and for the new impact parameter.

As the antiproton slowed, $b_{max}$ was increased in stages: we began with
$b_{max} = 1.0$; when $\pb$ energies dropped below 2.3 a.u., the level at
which small numbers of captures first began to be observed, $b_{max}$ was
increased to 2.0; when $\pb$ energies dropped below 1.2, somewhat above the
range where capture cross sections began to flatten out with $b_{max}$ of
2.0, $b_{max}$ was increased to 3.0.

While somewhat artificial, by selecting for collisions during which
something happened this method allowed us to study a realistic distribution
of antiproton projectiles during the last stages of antiproton cascade down
to eventual capture, typically after five to twenty sequential collisions,
while decreasing the amount of computer time spent modeling weakly
interacting collisions of large energy and impact parameter.

A total of 4,000 collision sequences (approximately 41000 individual He-
antiproton collisions) were followed.  916, or about $23\%$, ended with
capture via single electron ionization into the neutral exotic $\a e\bar p$,
while the remaining $77\%$ resulted in capture via double ionization into a
positively charged $\a\bar p$

Fig. 2a shows our total energy loss cross sections, $d(\si \Delta E) /dE,$
of the $\bar p$ as a function of initial $\bar p$ energy.  In the higher
energy collisions at the start of a collision sequence, little slowing of
the antiprotons occurs for large impact parameters.  As the antiprotons
slow, energy loss increases at larger impact parameters until the
antiprotons drop into the energy range from which capture begins to occur,
slightly below the energy at which we first increased the maximum impact
parameter; step increases in these energy loss cross sections are clearly
visible at energies of 2.3 a.u. and 1.2 a.u., where we increased the maximum
impact parameter.

Fig. 2b shows our antiproton capture cross sections for these collisions via
single and double ionization, again as a function of antiproton energy.  As
the antiprotons slow, the maximum impact parameter at which capture occurs
grows; since the central Coulomb potential of the He nucleus has the form
1/r, the total capture cross-section becomes infinite for very low energies.
Here, our total capture cross section levels out to $\pi b_{max}^2$ at the
lowest antiproton energies, where all the antiprotons are captured; our
maximum impact parameter of 3.0 for these lower energy collisions was chosen
so that this saturation occurred only for $\pb$ energies of approximately
0.1 a.u. or lower.

Due to the limited amount of energy and angular momentum that can be exchanged
by the massive antiproton and the bound electrons of the He atom, antiproton
energies and angular momenta which result in capture fall into a fairly narrow
range, as shown in Figs. 3a and 3b.  Fig. 3a shows the range of incident
antiproton energies and angular momenta which result in capture via double
electron ionization; the adjacent range of higher energy and angular momenta
antiprotons which were captured via single ionization is shown in Fig. 3b.

Fig. 4a shows the energy and angular momentum of the final states into which
the antiproton was captured via double ionization.  For capture via single
ionization, Fig. 4b shows the $\a\bar p e$ system states; in Fig. 4c, just
the antiproton states of these systems are plotted.  Here, after capture the
antiproton continues to interact and exchange energy and angular momentum
with the remaining electron, so each of states plotted in Fig. 4c represents
a time average of the antiproton configurations.  Fig. 5a,b, showing the
changing electron and antiproton radii as they interact over time, is an
example of a typical $\a e \pb$ system dynamics after antiproton capture.

Density contours of these scatter plots, in percent of total collisions per
unit E per unit $L$, are combined in Fig. 6, illustrating how the adjacent
bands of incident antiprotons of Fig. 3 drop into the captured states of
Fig. 4, and the relationship of the states of the antiprotons captured via
single ionization to the states of the total system.

Also of interest are the transition $E$ and $L$ distributions of the
antiprotons when they are first captured by the He atom.  Fig. 7a shows the
initial capture states of the antiprotons in what will become the doubly
ionized $\a \pb$, when the system was an unstable $\a e\pb $ or, in some
cases, $\a ee\pb $.  Fig. 7b shows the higher angular momentum states of
the antiprotons captured into what will become the singly ionized $\a e
\pb$, again at the point of initial antiproton capture.  Comparison with
Figs. 3 and 4 shows the incident antiprotons with higher angular momentum of
Fig. 3b clearly separating into the higher angular momentum states
associated with the more stable $\a e\pb $ of Fig. 4b, relative to the lower
angular momentum antiprotons in those systems that will quickly ionize to
the $\a\pb$ states of Fig. 4a.

\section{CAPTURE ANALYSIS}
\label{sec:anal}

The quantum antiproton levels for the simple $\a\pb$ are given above by
Eq.~\ref{eq:he}.

Since $n_0 \approx 38$ has generally been considered the lower limit of the
Auger cascade, i.e. the level below which $\pb$ orbits are progressively
less disturbed by any electrons which may still be attached, this has
been assumed to be the boundary between Auger and radiative decay, where
exotic projectiles are expected to accumulate following the Auger cascade
before further decay by radiative processes; much below this level decay
proceeds primarily by radiative capture irrespective of the state of
ionization.

In our calculations, antiprotons captured into $\a\pb$ are distributed in
this region; Fig. 4a shows the antiprotons stacked up against the classical
$L= n$ centrifugal boundary in the region of $n\approx 30-40$, with lower
$L$ states more heavily populated in the region $n< 35$.  It is not
surprising that low energy antiprotons are captured into states of $n < n_0$
via double ionization, since the antiproton must exchange enough energy with
the atomic electrons to boost both of them into (positive energy) escape
from the He nucleus.

It is also generally assumed that the final populations of these states are
distributed according to the statistical factor $2\ell+1$.  Fig. 8a shows
the distribution of the $\a\pb$ states vs. $L/n$, where the classical
circular angular momentum is given by

\be L_c = n = \sqrt{2 M^*/|E|}   \eq
In this figure, the total $L$-state distribution, averaged over the
different $n$-states, is approximately linear in $L$.

The separate higher energy and angular momentum band of $\a e \pb $ system
states shown in Fig. 4b can be compared with the $\a e \pb $ system level
diagrams proposed by Yamazaki {\it et al.},\cite{yam}, shown in Fig. 1; as
discussed above, some of the higher $L$ system states resulting from
additional angular momentum contributed by the remaining electron in the
singly ionized systems are indeed populated in our calculations.

The distinct, narrow band of high $L$ excited $\pb$ states present in the
singly ionized system (Fig. 4c) stack up against a modified centrifugal
barrier, at a lower value of $L$ than in the doubly ionized system due to
the screening effect of the remaining electron.  The degree to which the
antiprotons stack up against this barrier is illustrated by the population
distribution of Fig. 8b.  Here, the population distribution is plotted
against $L / n_{eff}$, where $n_{eff}$, labeling the energy levels of the
screened system,
\be E_n = - {Z_{eff}^2 M^*\over 2 n^2}, \eq
is given by
\be n_{eff} = \sqrt{Z_{eff}^2 M^* / E}, \eq
with $Z_{eff},$ defined in terms of the central radial force felt by the
antiproton
\be F_r = -Z_{eff} / r^2. \eq

\section{METASTABILITY}
\label{sec:meta}

As discussed above, it is a small percentage of antiprotons in stable, high
$L$ ``circular" states, from which the antiproton cannot decay to the point at
which it overlaps the nucleus and annihilation occurs, that has been
proposed as the source of the observed metastability in long-lived exotic
$\a e\pb $ atom \cite{iwa}.   In our calculations, a relatively large number
of the $\a e\pb$ atoms are in these higher $L$ states; it is, in fact, the
higher initial angular momentum of the incident antiprotons which results in
their capture via single ionization, as can be seen by comparing the initial
antiproton angular momenta, at the start of the collisions resulting in
antiproton capture, of Figs. 3a (double ionization) and 3b (single) with the
final system angular momentum distributions shown in Figs. 4a and b.

The question of what constitutes a stable, singly-ionized $\a e\pb$ system
is of course related to how long the system is followed. In our initial
calculations, systems were followed for a maximum time of 25,000 a.u.,
primarily due to limits in computational resources.  Indeed, some small
fraction of the systems classified as singly ionized have very small binding
energy, as can be seen in Fig. 4c.  In most of these cases where the $\pb$
is very loosely bound, the $\pb$ will eventually escape from its exotic
system, to be captured by the next He atom it encounters.

To further investigate $\a e\pb$ stability in our model, 500 systems were
followed to a maximum time of $10^5$ a.u.  As shown in Fig. 9, the
percentage of singly ionized systems surviving on these time scales is well-
described by a double exponential of the form

\be P_{\a e \pb}(t) = P_0 + P_1 e^{-t/\tau_1} + P_2 e^{-t/\tau_2}
                    = \left[22.3 + 82.6 e^{-t/2051} + 7.67 e^{-t/57,650}
                       \right]\%
\eq
This fit gives a survival of 22.3\% up to time $10^5$ a.u.
compared with the observed
metastable population of 3.6\% for survival beyond about $1\mu$s $\approx 4
\ 10^{10}$ a.u.
Some decay in the singly ionized population is observed at the end of this
time scale, suggesting the existence of
additional, longer time scale decay constants for this population.

\section{Conclusions and Future}
\label{sec:conc}

Semi-classical Monte Carlo calculations of the collisions of antiprotons
with He offer new details of the slowing and capture processes, including
the dependence of the final products on the incoming energy and angular
momentum and the state distributions of the singly and doubly ionized exotic
$\a e \pb $ and $\a\pb$ at the end of the Auger cascade.  These details shed
light on the recently reported metastable states of $\a e \pb  $.

Further questions to be investigated with this method include the effects
of impurities in the target medium on the slowing and capture process, and
more detailed study of the Auger process from which the high L meta-stable
states arise.  The population decay of these systems near the end of our
time scale invites further study of the dynamics of this decay, and of
their relation to the much longer time scales of the metastability discussed
by Yamazaki {\it et al.},\cite{yam}, who quote mean lifetimes of about
3 $\mu$s $\approx 1.25\ 10^{11}$ a.u., which is beyond the range of our
current computational tools.

We can suggest some mechanisms which might reduce survival to 1$\mu$s (which
we note is 4 10$^5$ times our longest runs.
One is Stark mixing into $s$-states due to the other helium atoms in
the liquid, leading to annihilation on the He nucleus.
This is inhibited by the one electron, but may still be
significant.  Another is annihilation from the higher $L$-states which we
find to survive.  We have
the population of these states.  These could be used by those who have
calculated the Born-Oppenheimer states to calculate the annihilation
rate from their  $\pb$ wave functions at the He nucleus.

\acknowledgments

The authors are grateful to Prof. Toshimitsu Yamazaki for discussions of his
work on this problem.  This work was supported in part by the DOE; one of us
(MAA) was supported in part by the NSF.

\newpage

FIG. 1.  The energy levels of of the $\a\,e\,\pb$ system courtesy of
Yamazaki and Ohtsuki \cite{yamoh}.
\vskip 12pt

FIG. 2\,a.  Energy loss cross sections for 41,417 $\pb$-He collisions,
as a function of initial $\pb$ kinetic energy, $T$.
\vskip 12pt

FIG. 2\,b.  Capture cross sections for 4,000 $\pb$-He collisions, as a function
of initial $\pb$ kinetic energy, $T$.  Capture into $\pb\,e\,\a$ states are
deno
   ted
by $(+)$, into $\pb\,\a$ by $(\times)$ and total by (*).
\vskip 12pt

FIG. 3\,a. Initial state of the $\pb$ for the last pass in the cascade
before capture into a $\pb\,\a$ state.
\vskip 12pt

FIG. 3\,b.  Initial state of the $\pb$ for the last pass in the cascade
before capture into a $\pb\,e\,\a$ state.
\vskip 12pt

FIG. 4\,a.  Final energy and angular momentum for 3,084 $\a\,\pb$ states.
\vskip 12pt

FIG. 4\,b.  Final total system energy and angular momentum for 916 $\a\,e\,\pb$
states.
\vskip 12pt

FIG. 4\,c.  Time averaged $\pb$ energy and angular momentum for 916
$\a\,e\,\pb$ states.
\vskip 12pt

FIG. 5\,a.  Radial coordinate of the $\pb$ as a function of time for a
particular run, $T_0=0.5\,,b_0=0.977$.
\vskip 12pt

FIG. 5\,b.  Radial coordinate of the electron as a function of time, in
comparison to a portion of the antiproton trajectory shown in Fig. 5\,a.
Because of the
rapid oscillation of the electron, the figure does not show all of the
calculated points.
\vskip 12pt

FIG. 6  Density contours, in percent per unit E per unit L.
The various regions (islands) describe:
A. The state of the incident antiprotons for
the final collision which ends in capture via double ionization.
B. The state of the incident antiprotons for
the final collision which ends in capture via single ionization.
C. Final (time-averaged) $\pb$ $E$ and $L$ in the $\a\,e\,\pb$
configuration
D. Final system states for the $\a\,\pb$ configuration, and
E. Final system states for the $\a\,e\,\pb$ configuration.
\vskip 12pt

FIG. 7\,a. Energy and angular momentum at initial $\pb$ capture for
collisions leading to $\a\,\pb$ products.
\vskip 12pt

FIG. 7\,b. Energy and angular momentum at initial $\pb$ capture for
collisions leading to $\a\,\e\,\pb$ products.
\vskip 12pt

FIG. 8\,a. Distribution in the ratio of angular momentum states $L$ to the
classical maximum angular momentum state $L_{max}=n$ for final $\a\,\pb$
states.  Averaging over the different $n$-states shows an approximately linear
distribution in the $L$-states populated.
\vskip 12pt

FIG. 8\,b. Distribution in the ratio of $\pb$ angular momentum states $L$
to the classical maximum angular momentum state $L_{max}=n_{eff}$ for final
$\a\,\e\,\pb$ states.  Averaging over the different $n$-states here shows
the concentration in high $L$ states of the $\pb$s trapped into
$\a\,\e\,\pb$ systems.
\vskip 12pt

FIG. 9. Survival as a function time of $\a\,\e\,\pb$ systems after initial
$\pb$ capture.


\begin{references}
\bibitem{iwa} M. Iwasaki {\it et al.}, Phys.\ Rev.\ Lett. {\bf 67}, 1246
(1991).
\bibitem{yam} T. Yamazaki {\it et al.},  Phys.\ Rev.\ Lett. {\bf 63}, 1590
(1989).
\bibitem{con} G. T. Condo, Phys.\ Lett.\ {\bf 9}, 65 (1964).
\bibitem{rus} J. E. Russell, Phys.\ Rev.\ Lett.\ {\bf 23}, 63 (1969);
 Phys.\ Rev.\ {\bf 188}, 187 (1969);
 Phys.\ Rev.\ A {\bf 1}, 721, 735, 742 (1970);
J.\ Math.\ Phys.\ {\bf 12}, 1906 (1971);
 Phys.\ Rev.\ {\bf A6}, 2488 (1972).
\bibitem{kir}   C. L. Kirschbaum and L. Wilets,
Phys.\ Rev.\ A {\bf 21}, 834 (1980).
\bibitem{yamoh} T. Yamazaki and K. Ohtsuki, Phys.\ Rev.\ A {\bf 45}, 7782
(1992).
\bibitem{wil2}  D. J. E. Callaway, L. Wilets, and Y. Yariv,
Nucl.\ Phys.\ A {\bf 327}, 250 (1979).
\bibitem{m&f}  P. M. Morse and H. Feshbach, Methods of Mathematical Physics,
p. 646, problem 5.3 (McGraw-Hill, New York, 1953)
\bibitem{shim}  I. Shimamura, Phys.\ Rev.\ A {\bf 46}, 3776 (1992).

\end{references}
\end{document}